\RequirePackage{ifpdf}

\documentclass[a4paper,11pt]{article}
\pdfoutput=1 % if your are submitting a pdflatex (i.e. if you have
\usepackage{jheppub} % for details on the use of the package, please
\usepackage[T1]{fontenc} % if needed

\usepackage{ifpdf} % to get it to work with pdflatex and not break for everything else

\usepackage{flexisym}
\usepackage{breqn}
\usepackage{comment}

\author{Costas G. Papadopoulos$^{a,b}$, Damiano Tommasini$^{a}$ and Christopher Wever$^{a}$}

\affiliation{$^{a}$Institute of Nuclear and Particle Physics, NCSR `Demokritos', Agia Paraskevi, 15310, Greece}
\affiliation{$^{b}$MTA-DE Particle Physics Group, University of Debrecen, H-4010 Debrecen, Hungary}

\emailAdd{costas.papadopoulos@cern.ch, tommasini@inp.demokritos.gr, wever@inp.demokritos.gr}


\title{Two-loop Master Integrals with the Simplified Differential Equations approach}

\abstract{We calculate the complete set of two-loop Master Integrals with two off mass-shell legs with massless internal propagators, that contribute to amplitudes 
of  diboson $V_1V_2$ production at the LHC. This is done with the Simplified Differential 
Equations approach to Master Integrals, which was recently proposed by one of the authors.}

\ifpdf
\fi

\begin{document}
\unitlength1cm
\maketitle
\allowdisplaybreaks

\section{Introduction}

At the first run of the LHC, proton collisions have reached the record-setting high energies of 8 TeV and delivered luminosity of $\sim23.3 \ \text{fb}^{-1}$. %recorded: ATLAS 21.7, CMS 21.79
At the second run of the LHC, which is expected to start in 2015, the energy and luminosity will be pushed much higher. In order to keep up with the increasing experimental accuracy as more data is collected, more precise theoretical predictions and higher loop calculations will be required.

With the better understanding of reduction of one-loop amplitudes to a set of Master Integrals ({\bf MI}) based on unitarity methods~\cite{Bern:1994cg,Bern:1994zx} and at the integrand level via the OPP 
method~\cite{Ossola:2006us,Ossola:2008xq}, one-loop calculations have been fully automated in many numerical tools (some reviews on the topic are~\cite{AlcarazMaestre:2012vp,Ellis:2011cr}).
%\cite{vanHameren:2009dr,Berger:2009zg,Bern:2013gka,Bredenstein:2009aj,Bevilacqua:2009zn,Bevilacqua:2010ve}\footnote{I don't like citations: GoSam,Mad-family-F.Maltoni, Feynart-T.Hahn, OpenLoops-S.Pozzorini...}.
In the recent years, a lot of progress has been made towards the extension of these reduction methods to the two-loop order at the integral~\cite{Gluza:2010ws,Kosower:2011ty,CaronHuot:2012ab} 
as well as the integrand~\cite{Mastrolia:2011pr,Badger:2012dp,Badger:2013gxa,Papadopoulos:2013hra} level. Contrary to the MI at one-loop, which have been known for a long time already~\cite{'tHooft:1978xw}, a complete library of MI at two-loops is still missing.
At the moment this seems to be the main obstacle to obtain a fully automated NNLO calculation framework similar to the one-loop case, that will satisfy the anticipated precision requirements at the LHC~\cite{Butterworth:2014efa}.

Starting from the works of~\cite{Goncharov:1998kja,Remiddi:1999ew,Goncharov:2001iea}, there has been a building consensus that the so-called {\it Goncharov Polylogarithms} ({\bf GPs}) form a functional basis for many MI. A very fruitful method for calculating MI and expressing them in terms of GPs is the differential equations ({\bf DE}) approach~\cite{Kotikov:1990kg,Kotikov:1991pm,Bern:1992em,Remiddi:1997ny,Henn:2013pwa}, which has been used in the past two decades to calculate various MI at two-loops ~\cite{Caffo:1998du,Gehrmann:1999as,Gehrmann:2000zt,Gehrmann:2001ck,Henn:2014lfa,Caola:2014lpa}.
%and more recently a new idea was proposed in~\cite{Henn:2013pwa}. 
In~\cite{Papadopoulos:2014lla} a variant of the traditional DE approach to MI was presented, which was coined the Simplified Differential Equations ({\bf SDE}) approach. In this paper a review of the SDE approach and an application to the four-point two-loop planar and non-planar MI with two different off-shell external legs and massless internal propagators is presented.

Those double-box integrals are needed in particular in order to compute NNLO QCD corrections to $pp\rightarrow VV^*$ processes at LHC, where $VV^*$ are massive electroweak vector bosons. In particular the case of two different masses can be helpful to consider $ZZ^*$ or $W^+W^-$ off-shell production as well as different bosons production like $WZ$ (or $V\gamma^*$, $ZH$,  $\gamma^*\gamma^*$, etc.). 
%Improving the theoretical results of diboson production could help to clarify discrepancies between measured and predicted kinematical distributions, in particular for the process $pp\rightarrow W^+W^-$~\cite{Chatrchyan:2013oev,ATLAS:2012mec}\footnote{not sure about these citations!!!}. 
Another significant improvement is related to the fact that $pp\rightarrow VV^*$ is the main contribution to the background for testing Higgs boson decay into two vector bosons. A better understanding of this background could lead to improved measurements of Higgs couplings to vector bosons as well as the Higgs decay width~\cite{Caola:2013yja,Campbell:2013una}.
%(that is basically measured by considering off-shell decay to $H\rightarrow ZZ^*$~\cite{Caola:2013yja,Campbell:2013una}).

For these reasons, recently two loop MI with two off-shell external legs have been intensively studied and calculated by different groups~\cite{Gehrmann:2013cxs,Gehrmann:2014bfa,Henn:2014lfa,Caola:2014lpa}. Furthermore those results have been already implemented to provide NNLO predictions for $ZZ$ and $W^+W^-$ production at the LHC~\cite{Cascioli:2014yka,Gehrmann:2014fva}. In this paper we calculate the same MI as those given in~\cite{Henn:2014lfa,Caola:2014lpa}, namely the full set of two loop MI with different masses which contains both planar as well as non-planar Feynman diagrams, by using the independent SDE approach presented in~\cite{Papadopoulos:2014lla}.

The paper is organized as follows. In the Section 2 we explain and review the SDE method. In Section 3 we give the parameterization and notation of the variables describing the two-loop MI. As we will show, the same parameterization of the external legs can be used to parametrize both the planar {\it and} non-planar MI. %In Section 4 we explain how to perform analytic continuation of results from Euclidean to physical solutions. 
In Section 4 we discuss our results, we show how these results can be analytically continued in order to accommodate both the Euclidean as well as the physical kinematical regions and compare with results obtained 
by direct numerical integration as well as with the analytic calculations given in~\cite{Henn:2014lfa}. 
%We also briefly discuss how to perform the analytic continuation of our results from the Euclidean region to the physical region. 
We conclude in Section 5 and provide an overview of the topic and some perspective for future developments. 
In the Appendix we give few characteristic examples on how the boundary conditions are properly reproduced in our approach by the DE. 
Finally in the attached ancillary files, we provide our analytic results for all  two-loop MI in terms of Goncharov polylogarithms together with explicit numerical results.

\section{Simplified differential equations}

\begin{figure}[t!]
\centering
\includegraphics[width=0.4 \linewidth]{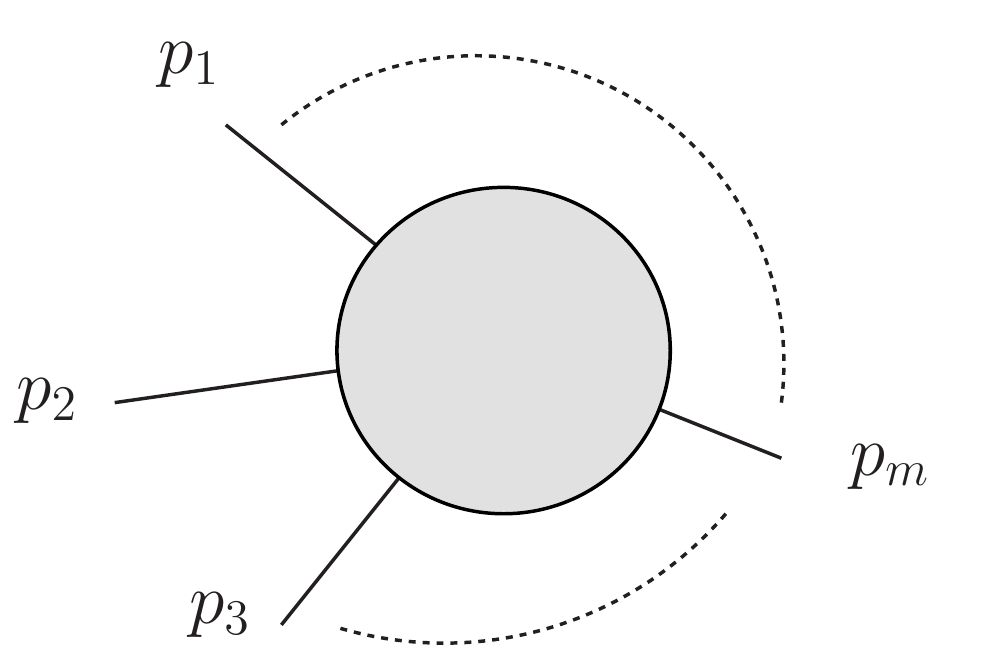} \hspace{1.5 cm}
\includegraphics[width=0.4 \linewidth]{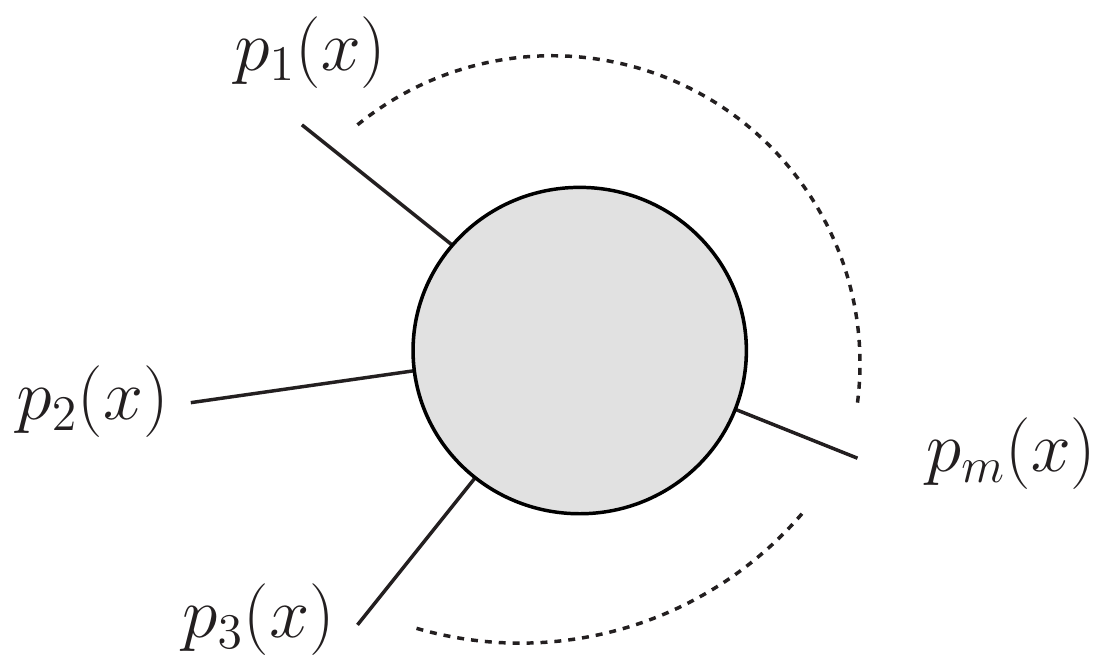}
  \caption{General graphs represented by (\protect\ref{eq:loopgen}) on the left and (\protect\ref{eq:loopgenx}) on the right.}
  \label{fig:xparam}
\end{figure}

The MI in this paper will be calculated with the SDE approach~\cite{Papadopoulos:2014lla} and therefore in this Section a brief review is given of how the method is used in practice%\footnote{We refer to~\cite{Papadoupoulos:2014dla} for another review of the (simplified) differential equations method.}
. Assume that one is interested in calculating an $l-$loop Feynman integral whose graph with external momenta $\{p_j\}$, considered incoming, is shown on the left hand side in Figure \ref{fig:xparam}. Throughout this paper the internal propagators are assumed massless.
%, since the method has been originally developed for massless internal lines. 
The Feynman integral on the left hand side in Figure \ref{fig:xparam}, is a subset of the following class of $l-$loop integrals:
\begin{equation}
G_{a_1\cdots a_n}(\{p_j\},\epsilon)=\int\left(\prod_{r=1}^l \frac{d^dk_r}{i\pi^{d/2}}\right)\frac{1}{D_1^{a_1}\cdots D_n^{a_n}}, \hspace{0.5 cm}
D_i=\left(c_{ij}k_j+d_{ij}p_j\right)^2,\,\, d=4-2\epsilon
\label{eq:loopgen}
\end{equation}
with matrices $\{c_{ij}\}$ and $\{d_{ij}\}$ determined by the topology and the momentum flow of the graph, and
the denominators are defined in such a way that all scalar product invariants can be written as a linear combination of them. The exponents $a_i$ are integers and may be negative in order to accommodate non-trivial numerators.

Any integral $G_{a_1\cdots a_n}$ may be written as a linear combination of the MI $\vec{G}^{MI}(\{ s_{ij} \},\epsilon)$ with coefficients depending on the independent scalar products, $s_{ij}=p_i\cdot p_j$, and space-time dimension $d$, by the use of {\it integration by parts} ({\bf IBP}) identities~\cite{Chetyrkin:1981qh,Tkachov:1981wb}.  In the traditional DE method, the vector of MI $\vec{G}^{MI}(\{s_{ij}\},\epsilon)$ is differentiated with respect to $p_i \cdot \frac{\partial}{\partial p_j}$ 
%a Lorentz invariant (or set of invariants) $\textbf{s}$
, and the resulting integrals are reduced by IBP to give a linear system of DE for 
$\vec{G}^{MI}(\{s_{ij}\},\epsilon)$~\cite{Kotikov:1990kg,Remiddi:1997ny}. The  invariants, $s_{ij}$, are then parametrised in terms of dimensionless variables, defined on a case by case basis, so that the resulting DE can be solved in terms of GPs. 
Usually boundary terms corresponding to the appropriate limits of the chosen parameters have to be calculated 
using for instance expansion by regions techniques~\cite{Beneke:1997zp,Smirnov:2002pj}.

In the SDE approach~\cite{Papadopoulos:2014lla} a dimensionless parameter $x$ is introduced so that the vector of MI $\vec{G}^{MI}(\{s_{ij}\},\epsilon;x)$, which now depends on $x$, satisfies
\begin{equation}
\frac{\partial}{\partial x} \vec{G}^{MI}(\{s_{ij}\},\epsilon;x)=\mathbf{H}(\{ s_{ij}\},\epsilon;x)\vec{G}^{MI}(\{s_{ij}\},\epsilon;x)\label{eq:DEx}
\end{equation}
a system of differential equations in one independent variable. The expected benefit is that the integration of the DE naturally captures the expressibility of MI in terms of GPs and more importantly makes the problem 
 {\it independent on the number of kinematical scales} (independent invariants) involved. 
 As shown on the right hand side in Figure \ref{fig:xparam}, the external incoming momenta are now {\it parametrized} linearly in terms of $x$ as $p_i(x)=p_i+(1-x)q_i$, where the $q_i$'s are a linear combination of the  momenta $\{p_i\}$ such that $\sum_iq_i=0$. If $p_i^2=0$, the parameter $x$ captures the off-shellness of the external legs. 
The class of Feynman integrals in (\ref{eq:loopgen}) are now dependent on $x$ through the external momenta:
\begin{equation}
G_{a_1\cdots a_n}(\{s_{ij}\},\epsilon;x)=\int\left(\prod_{r=1}^l \frac{d^dk_r}{i\pi^{d/2}}\right)\frac{1}{D_1^{a_1}\cdots D_n^{a_n}}, 
\;\;\;
D_i=\left( c_{ij}k_j+d_{ij}p_j(x) \right)^2.
%\hspace{0.5 cm} s=\{p_i.p_j\}|_{i,j}, 
\label{eq:loopgenx}
\end{equation}
%where contrary to the traditional DE approach, 
%where the Lorentz invariants $s$ are here defined as the usual scalar products. 
Note that as $x\rightarrow 1$, the original configuration of the loop integrals (\ref{eq:loopgen}) is reproduced, which correspond to a simpler one with one scale less. 
%The vector 
%$\vec{G}^{MI}(x)$ is now dependent on $x$ and one differentiates it with respect to $x$ to get a linear system of 
%differential equations:
%\begin{equation}
%\text{SDE:} \hspace{1 cm} \frac{\partial}{\partial x} \vec{G}^{MI}(x,s,\epsilon)
%\stackrel{IBP}{=}\overline{\overline{H}}(x,s,\epsilon)\cdot\vec{G}^{MI}(x,s,\epsilon), \ \ \ \ s=\{p_i.p_j\}|_{i,j}. \label{eq:DEx}
%\end{equation}

We briefly summarize how the DE (\ref{eq:DEx}) may be solved in practice. The MI with least amount of denominators $m_0$ ($m_0=3$ in the case of two loops) are 
typically two-point 
integrals 
that satisfy homogeneous equations and  can be easily calculated analytically by other methods. Furthermore, the form of \ref{eq:DEx} is such that MI with $m$ denominators  only depend on MI with at most $m$ denominators. This structure of the DE makes it possible to first solve the MI with $m_0+1$ denominators, then those with $m_0+2$ denominators and so forth. In other words, in practice the DE may be solved in a {\it bottom-up} approach. 

Assume that all MI with $m'\leq m$ denominators are known and already expressed in the desired form, the meaning of which will become clear below. The DE of MI with $m+1$ denominators can be written schematically in the form:
\begin{equation}
\partial_x G_{m+1}=H\left(\{s_{ij}\},\epsilon;x \right)G_{m+1}+\sum_{m'\geq m_0}^{m} R\left(\{s_{ij}\},\epsilon;x \right) G_{m'}, \label{eq:DEm}
%\partial_x G(x,s,\epsilon)=H(x,s,\epsilon)G(x,s,\epsilon)+\sum_{n,l} x^{-n+l\epsilon}I_{n,l}(x,s,\epsilon). \label{eq:DEm}
\end{equation}
i.e. the sum of a homogeneous and an inhomogeneous term. The functions $H$ and $R$ are rational functions of their arguments.
%with $n\in \mathbb{Z}, n\geq 0$ and $l\in \mathbb{Q}$.  
Equation (\ref{eq:DEm}) can be  solved with the variation of constants method by introducing the integrating factor $M\left(\{s_{ij}\},\epsilon;x \right)$ which satisfies the differential equation 
$\partial_x M=-MH$ (dropping the arguments of the functions for brevity):
\begin{equation}
\partial_x (MG_{m+1})=M\sum_{m'\geq m_0}^{m}  R\left(\{s_{ij}\},\epsilon;x \right) G_{m'}.
%=\sum_{n,l} x^{-n+l\epsilon}MI_{n,l}(x,s,\epsilon)=:\sum_{n,l} x^{-n+l\epsilon}\tilde{I}_{n,l}(s,\epsilon)+\sum_lx^{l\epsilon}\tilde{I}_{0,l}(x,s,\epsilon). 
\label{eq:DEmint}
\end{equation}
The equation could now be straightforwardly  expanded in $\epsilon$ and integrated, provided that the right-hand side is free of singularities at $x\to 0$. In order to deal with possible singularities at $x=0$ we need 
 therefore to determine the re-summed part of the solution in this limit. 

The righthand side of (\ref{eq:DEmint}) can be rewritten schematically as a sum of a {\it singular} and {\it regular} term at $x=0$:
\begin{equation}
M\sum_{m'\geq m_0}^{m}  R\left(\{s_{ij}\},\epsilon;x \right) G_{m'}=: \sum_i x^{-1+\beta_i \epsilon}\tilde{I}^{(i)}_{sin}(\{s_{ij}\},\epsilon)+\tilde{I}_{reg}(\{s_{ij}\},\epsilon;x). 
\label{eq:DEsing}
\end{equation}
with $\beta_i$ being typically rational numbers.
The singular term is integrated exactly and the solution of (\ref{eq:DEmint}) becomes:
\begin{equation}
MG_{m+1}=C(\{s_{ij}\},\epsilon)+ \sum_i \frac{x^{\beta_i \epsilon}}{\beta_i \epsilon}\tilde{I}^{(i)}_{sin}(\{s_{ij}\},\epsilon)+ \int_0^x dx' \tilde{I}_{reg}(\{s_{ij}\},\epsilon;x'), 
\label{eq:DEfin}
\end{equation}
where the first term $C(\{s_{ij}\},\epsilon)$ is a constant in $x$ but may be dependent on the the kinematical invariants.  
The rightmost term in Eq.(\ref{eq:DEfin}), is safely expanded  in $\epsilon$ and expressed in terms of GPs. 
%In writing Equation (\ref{eq:DEmint}) 
To this end the integrating factors $M$ in Eq.(\ref{eq:DEmint}) 
should be rational functions of $x$ in the limit $\epsilon\to 0$. This is a {\it sufficient condition} for the chosen $x$-parametrisation to result in a differential equation
solvable in terms of GPs. 

As was noted in~\cite{Papadopoulos:2014lla}, in the SDE approach the {\it boundary terms},
%when solving from the bottom-up 
in all cases studied until now including one and two loops MI, are always naturally captured by the integrated singularities in the SDE (\ref{eq:DEfin}) themselves at $x=0$, which is precisely the lower integration boundary of the GPs.
%\footnote{In all cases we have considered $n=0,1$.}. 
In other words, in all cases studied the integrated singular terms in (\ref{eq:DEfin}) correctly describe the behaviour of $MG_{m+1}$ as $x\rightarrow 0$ and thus the constant $C(\{s_{ij}\},\epsilon)$ vanishes. In this way, the SDE method is well suited for directly and efficiently expressing the MI in terms of GPs without the need for an independent evaluation of the MI at the boundary $x=0$. 
In fact all re-summed parts in the limit $x\to 0$ are fully determined by the one-scale MI involved in the system --  in the case of the present calculation, two-point integrals, two three-point integrals and double one-loop integrals -- that satisfy homogenous differential equations.
These integrals are simple enough to be evaluated by other methods and in fact they are all known for the cases of interest.
As we have verified using the expansion by regions technique this property is closely related to the $x$-parametrization that allows the expression of the limit $x\to 0$ of all integrals involved in terms of the
one-scale integrals, see Appendix \ref{expbyreg}.   

%In particular, the boundary terms of all the two-loop graphs calculated in this paper as well as those for many other examples we have considered are captured by the SDE without the addition of any constants $C(s,\epsilon)$ in $x$ on the right hand side of (\ref{eq:DEfin}). This is of course a welcome {\it bonus}, tested in all cases {\it a posteriori}, but the method is still applicable when $C(s,\epsilon)$ has to be derived through other methods, as for instance by using the expansion by regions technique~\cite{Beneke:1997zp,Smirnov:2002pj}, as it is also the case in the traditional DE approach.

\begin{figure}[t!]
\centering
\includegraphics[width=0.25 \linewidth]{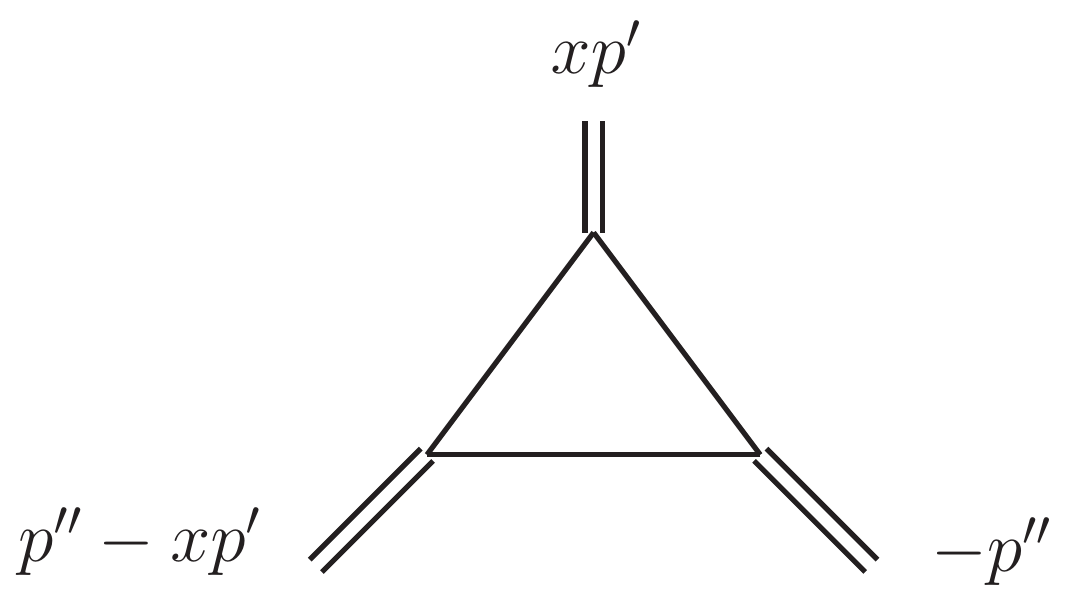}
  \caption{Required parametrization for off mass-shell triangles after possible pinching of internal line(s).}
  \label{fig:xparam-tr}
\end{figure}

The integrand of the remaining integral
%\footnote{Note that these are effectively just the usual plus distributions resulting from subtracting the singular part of the integrand.} 
in (\ref{eq:DEfin}) is regular at the boundary $x=0$.
% and therefore may be expanded in a series in $\epsilon$. 
After expanding in $\epsilon$ and performing partial fraction decomposition in $x$, 
%For a large class of MI 
the integral is directly expressible in terms of Goncharov polylogarithms, which are an iterative class of functions that generalize the usual logarithms and polylogarithms~\cite{Goncharov:1998kja,Goncharov:2001iea}:
\begin{gather}
GP(\underbrace{\alpha_1,\cdots ,\alpha_n}_{\text{weight n}};x):=\int_0^x dx' \frac{GP(\alpha_2,\cdots ,\alpha_n;x')}{x'-\alpha_1}, \nonumber\\
GP(;x)=1, \ \ \ \ GP(\underbrace{0,\cdots ,0}_{\text{n times}};x)=\frac{1}{n!}\log^n(x), \label{eq:GPs}
\end{gather}
where in general $\alpha_i,x\in \mathbb{C}$. Note that the integration boundary in (\ref{eq:DEfin}) was chosen to be $x=0$ precisely in order to directly express the integrals in terms of GPs. If another boundary point $x=x_0$ is chosen, the integrals in (\ref{eq:DEfin}) result in slightly more complicated expressions made up of differences of GPs. We will denote the first set of arguments of the GPs, i.e. those contributing to the weight, as {\it indices}. The SDE method therefore directly expresses the MI in terms of a well defined functional basis of GPs, {\it independently on the number of kinematical scales involved}.
%\footnote{In~\cite{Gehrmann:2014bfa,Henn:2014lfa,Caola:2014lpa} DE of the MI w.r.t. {\it multiple} variables are considered and the {\it letters} are defined through the total differential of the MI. In this paper the MI are only differentiated w.r.t. the variable $x$ and therefore the differential of the MI directly corresponds to the arguments of the GPs in the MI.}

When the DE are coupled, the homogeneous factor $\mathbf{H}$ is a {\it matrix} and $\vec{G}_{m+1}$ is a {\it vector} in (\ref{eq:DEm}). For all cases considered so far, proceeding the same way as in the uncoupled case, the diagonal elements are used to determine the integrating factor matrix, namely the diagonal matrix $\mathbf{M}_D$  satisfying 
$\partial_x \mathbf{M}_D=- \mathbf{M}_D  \mathbf{H}_D$, where $ \mathbf{H}_D$ is the diagonal part of $ \mathbf{H}$.
 
%, one may expand the DE in $\epsilon$ to get a power series. 
%For the two-loop double boxes considered in this paper 
%We find that the $\epsilon^0$ term  is a sum of a {\it triangular matrix} plus a diagonal piece. 
%After the diagonal terms are absorbed as integrating factors, 
The  homogeneous matrix $\tilde{\mathbf{H}}=: \mathbf{M}_D\left(  \mathbf{H}- \mathbf{H}_D\right) \mathbf{M}^{-1}_D$ of the reduced system of DE is then a {\it strictly triangular matrix} at order $\epsilon^0$ 
and the system becomes effectively uncoupled, 
 and may be easily integrated order by order in $\epsilon$ as was for example explained in~\cite{Papadopoulos:2014lla}.
Furthermore, singularities at $x=0$ in the inhomogeneous terms are integrated to determine the re-summed part of the solutions in exactly 
the same way as in the uncoupled DE case. 
One may also envisage to solve the reduced system by using a Magnus or Dyson series expansion~\cite{Argeri:2014qva,DiVita:2014pza}. 
 The latter approach is possible because the non-diagonal piece of the $\epsilon^0$ matrix is stricly triangular and thus nilpotent. We refer to~\cite{Gehrmann:2014bfa} for a similar discussion for the case of coupled DE.

In very few specific cases terms of the form $x^{-2+\beta_i \epsilon}$ appears in the matrix $\tilde{ \mathbf{H}}$, the homogeneous part of the reduced system of coupled DE
after the integrating factors have been taken into account. In these cases, by using the transformation $x\to 1/x$ we are able to bring the DE in a form with all singularities 
at $x=0$ being of the form $x^{-1+\beta_i \epsilon}$ and contained in the {\it inhomogeneous part} of the DE. 
We then follow the same procedure as in the case of uncoupled DE equations.
% (\ref{eq:DEsing}) and (\ref{eq:DEfin}). 
In fact as can be seen in the Appendix \ref{expbyreg}, the DE after the transformation do correctly reproduce the
boundary conditions in terms of single-scale integrals as usual. 
Of course the results are expressed back in terms of $x$ by applying the  $x\to 1/x$ transformation on the argument of the GPs. 

%For many cases, the MI with $m_0$ denominators are expressible in terms of GPs and one needs to choose the {\it parametrization} of the external momenta in $x$ such that the integrals in (\ref{eq:DEfin}) result in GPs (\ref{eq:GPs}). 
As far as the {\it parametrization} of the external momenta in $x$  is concerned, we noticed that for all the cases that we considered it was enough for us to choose the parametrization of the external legs such that after pinching internal lines the resulting triangles with three off-shell legs, if they appear, have the form given in Figure \ref{fig:xparam-tr}~\cite{Papadopoulos:2014lla}. 
%In other words one of the external momentum should scale linearly with $x$, and another one should be independent of $x$. 
%In the next Section we present the $x$-parametrization of the two-loop double boxes which satisfies this constraint and discuss our notation of the variables used for expressing our MI.

\section{Notation and $x$-parametrization of the Master Integrals}

\begin{figure}[t!]
\centering
\includegraphics[width=0.29 \linewidth]{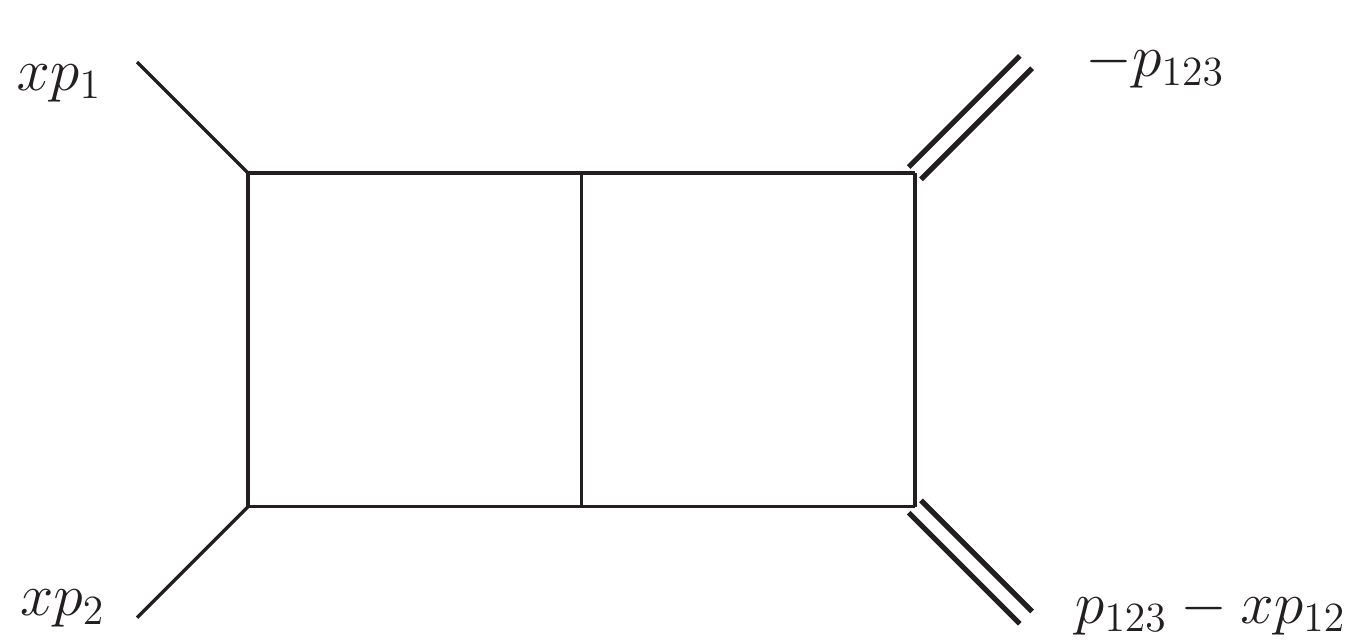} \hspace{0.4 cm}
\includegraphics[width=0.31 \linewidth]{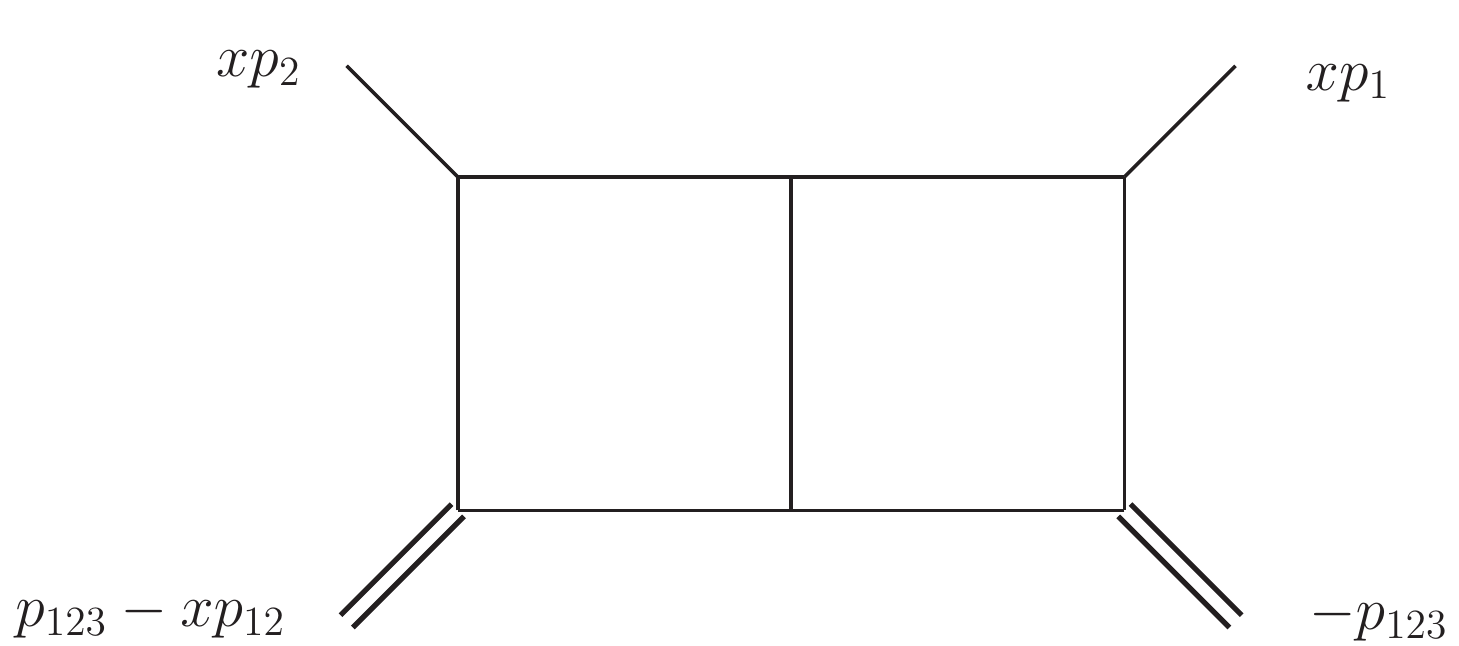} \hspace{0.4 cm}
\includegraphics[width=0.31 \linewidth]{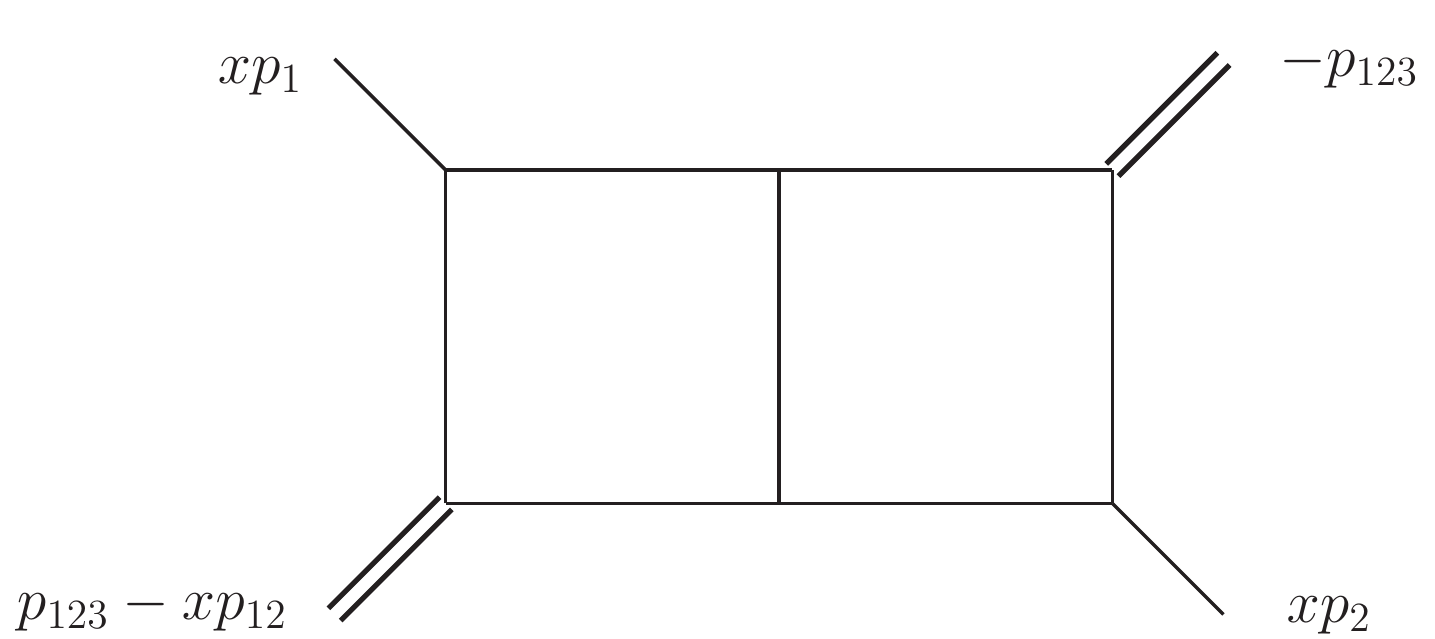}
  \caption{The parametrization of external momenta for the three planar double boxes of the families $P_{12}$ (left), $P_{13}$ (middle) and $P_{23}$ (right) contributing to pair production at the LHC. All external momenta are incoming.}
  \label{fig:xparam-P}
\end{figure}

\begin{figure}[t!]
\centering
\includegraphics[width=0.29 \linewidth]{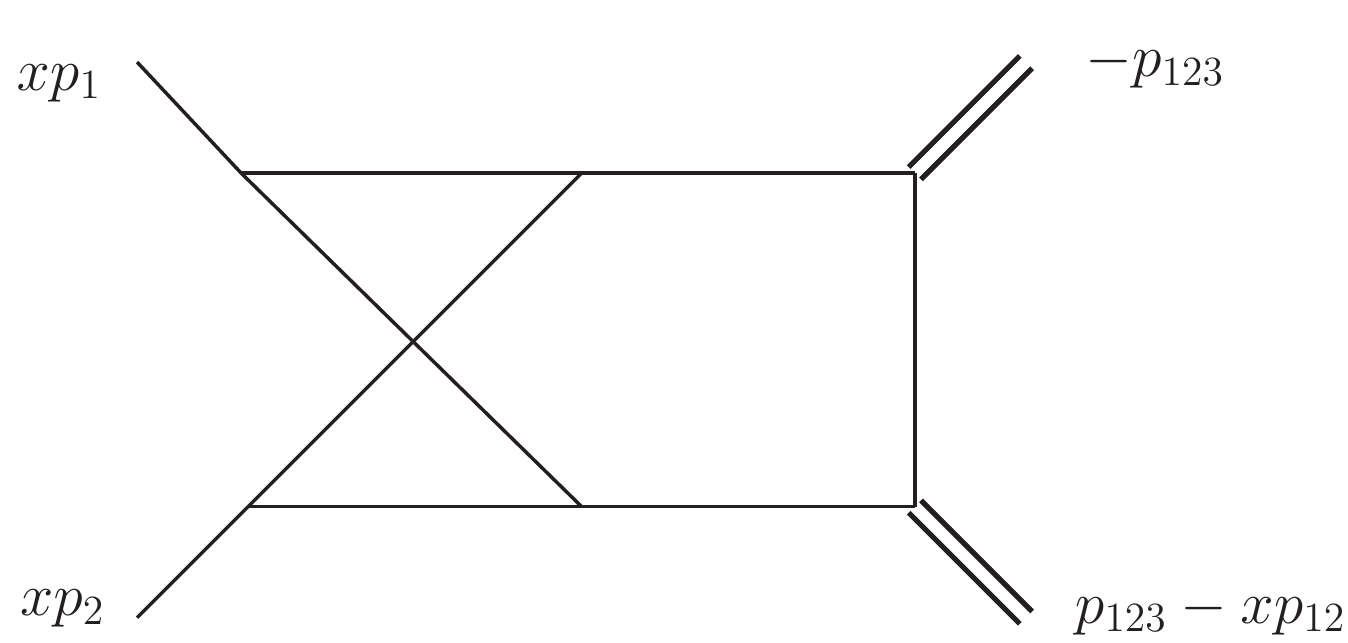} \hspace{0.4 cm}
\includegraphics[width=0.31 \linewidth]{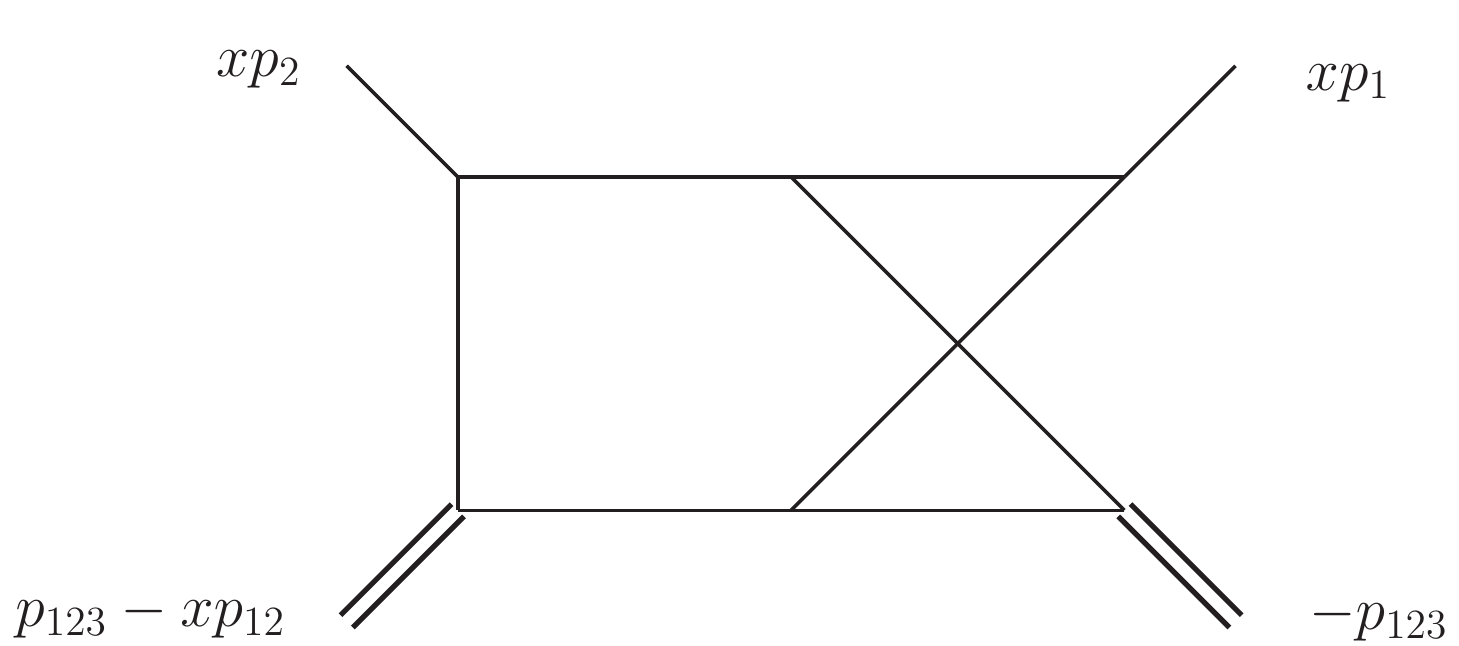} \hspace{0.4 cm}
\includegraphics[width=0.31 \linewidth]{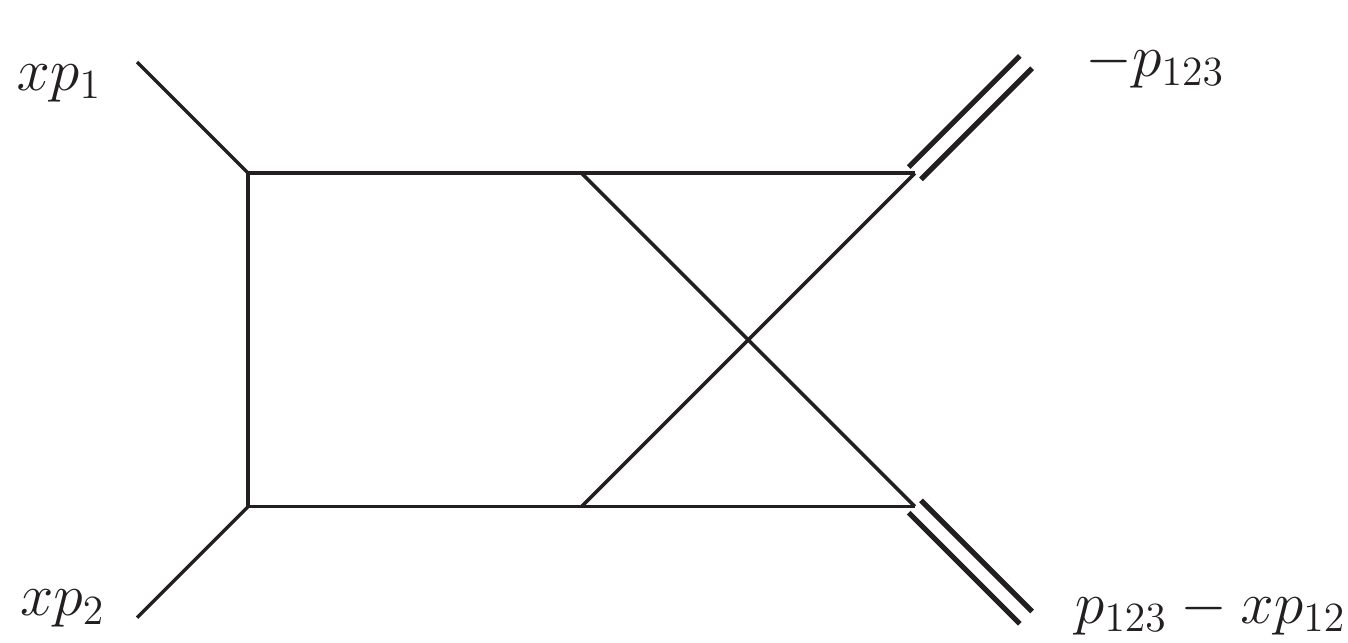}
  \caption{The parametrization of external momenta for the three non-planar double boxes of the families $N_{12}$ (left), $N_{13}$ (middle) and $N_{34}$ (right) contributing to pair production at the LHC. All external momenta are incoming.}
  \label{fig:xparam-N}
\end{figure}

We are interested in calculating the MI of two-loop QCD amplitude corrections contributing to diboson pair production at the LHC, where the two outgoing particles may have different masses:
\begin{gather}
p(q_1)p'(q_2)\rightarrow V_1(q_3)V_2(q_4), \ \ q_1^2=q_2^2=0, \ \ q_3^2=M_3^2, \ \ q_4^2=M_4^2.\label{eq:ppVV}
\end{gather}
The colliding partons have massless momenta $q_1,q_2$ and the outgoing particles have massive momenta $q_3,q_4$. Both the planar and non-planar diagrams contributing to (\ref{eq:ppVV}) have already been calculated with the traditional DE method~\cite{Henn:2014lfa,Caola:2014lpa}. There are in total six families of MI whose members with the maximum amount of denominators are graphically shown in Figures \ref{fig:xparam-P} and \ref{fig:xparam-N}. Three of these, namely those shown in Figure \ref{fig:xparam-P}, contain only planar MI and will therefore be referred to as the planar families in this paper. They will be denoted by $P_{12},P_{13}$ and $P_{23}$ as was done in~\cite{Henn:2014lfa} and contain 31, 29 and 28 MI respectively. The other three families, shown in Figure \ref{fig:xparam-N}, contain planar MI with up to 6 denominators as well as non-planar MI with 6 and 7 denominators and will be referred to as the non-planar families of MI. These non-planar families will be denoted by $N_{12},N_{13}$ and $N_{34}$ as was done in~\cite{Caola:2014lpa} and contain 35, 43 and 51 MI respectively. In this paper we have used both {\bf FIRE}~\cite{Smirnov:2008iw} and {\bf Reduze 2}~\cite{vonManteuffel:2012np} to perform the IBP reduction to the MI.

In order to parametrize the external momenta in terms of $x$ as was dicussed in the previous Section we kept in mind Figure \ref{fig:xparam-tr}. In particular, the parametrization of $P_{12}$ and $P_{13}$ were chosen such as to satisfy the requirement of having off-shell  triangles of the form shown in Figure \ref{fig:xparam-tr} after pinching the internal line(s) between the two massless legs in Figures \ref{fig:xparam-P}. The parametrization of $P_{23}$ was found by permuting the external legs accordingly\footnote{Note that for the family $P_{23}$ it is not possible to get off-shell triangles by pinching the internal lines.}. All non-planar diagrams shown in Figures \ref{fig:xparam-N} may be transformed, after pinching enough internal lines, to off-shell triangles with the exact same external momenta as those off-shell triangles found after pinching the families $P_{12}$ and $P_{13}$. Therefore in the SDE approach it is sufficient to choose the {\it same} parameterization of the external legs for the non-planar families as was chosen for the planar ones in order for the families to contain off-shell triangles of the form shown in Figure \ref{fig:xparam-tr}. The incoming external momenta for {\it all families} are therefore parametrized as follows:
\begin{gather}
q_1=xp_1, \ \ q_2=xp_2, \ \ q_3=p_{123}-xp_{12}, \ \ q_4=-p_{123}, \ \ p_i^2=0, \nonumber\\
s_{12}:=p_{12}^2, \hspace{0.5 cm} s_{23}:=p_{23}^2, \hspace{0.5 cm} q:=p_{123}^2, \label{eq:qxpar}
\end{gather}
where the notation $p_{i\cdots j}=p_i+\cdots +p_j$ is used. As $x\rightarrow 1$, the external momentum $q_3$ becomes massless, such that (\ref{eq:qxpar}) also captures the case where only one outgoing particle is off-shell. 
%The other scenario where both particles $V_1,V_2$ are massless can then be found by setting the variable $q=0$. 
The MI depend in total on 4 variables, namely the traditional Mandelstam variables $S=(q_1+q_2)^2,T=(q_1+q_3)^2$ and (squared) particle masses $M_3^2=q_3^2,M_4^2=q_4^2$. The $x$-parameterization (\ref{eq:qxpar}) of the external momenta results in these variables being related to the parameter $x$, the external off-shell mass squared  $q$ and the two independent scalar products $s_{12}$ and $s_{23}$ of our choice that were defined in (\ref{eq:qxpar}):
\begin{gather}
S=s_{12}x^2, \hspace{0.5 cm} T=q - (s_{12} + s_{23}) x, \hspace{0.5 cm} M_3^2=(1 - x) (q - s_{12} x), \hspace{0.5 cm} M_4^2=q. \label{eq:Pinv}
\end{gather}
The other Mandelstam variable $U=(q_1+q_4)^2$ is related to the other variables as usual via $S+T+U=M_3^2+M_4^2$. 

After finding the $x$-parameterization (\ref{eq:qxpar}), the class of loop integrals describing the planar families $P_{12}$, $P_{13}$ and $P_{23}$ are now explicitly expressed in $x$ as:
\begin{eqnarray}
G^{P_{12}}_{a_1\cdots a_9}(x,s,\epsilon)&:=e^{2\gamma_E \epsilon} &\int \frac{d^dk_1}{i\pi^{d/2}}\frac{d^dk_2}{i\pi^{d/2}}
\frac{1}{k_1^{2a_1} (k_1 + x p_1)^{2a_2} (k_1 + x p_{12})^{2a_3} (k_1 + p_{123})^{2a_4}} \nonumber\\
&\times& \frac{1}{k_2^{2a_5} (k_2 - x p_1)^{2a_6} (k_2 - x p_{12})^{2a_7} 
(k_2 - p_{123})^{2a_8} (k_1 + k_2)^{2a_9}}, \label{eq:P12x}
\end{eqnarray}
\begin{eqnarray}
G^{P_{13}}_{a_1\cdots a_9}(x,s,\epsilon)&:=e^{2\gamma_E \epsilon} &\int \frac{d^dk_1}{i\pi^{d/2}}\frac{d^dk_2}{i\pi^{d/2}}
\frac{1}{k_1^{2a_1} (k_1 + x p_1)^{2a_2} (k_1 + x p_{12})^{2a_3} (k_1 + p_{123})^{2a_4}} \nonumber\\
&\times& \frac{1}{k_2^{2a_5} (k_2 - x p_1)^{2a_6} (k_2 - p_{12})^{2a_7} 
(k_2 - p_{123})^{2a_8} (k_1 + k_2)^{2a_9}}, \label{eq:P13x}
\end{eqnarray}
\begin{eqnarray}
G^{P_{23}}_{a_1\cdots a_9}(x,s,\epsilon)&:=e^{2\gamma_E \epsilon} &\hspace{-0.1cm}\int \frac{d^dk_1}{i\pi^{d/2}}\frac{d^dk_2}{i\pi^{d/2}}
\frac{1}{k_1^{2a_1} (k_1 + x p_1)^{2a_2} (k_1 + p_{123}-x p_2)^{2a_3}
 (k_1 + p_{123})^{2a_4}} \nonumber\\
&\times& \hspace{-0.3cm}\frac{1}{k_2^{2a_5} (k_2 - p_1)^{2a_6} (k_2 + x p_2 - p_{123})^{2a_7} 
(k_2 - p_{123})^{2a_8} (k_1 + k_2)^{2a_9}}, \label{eq:P23x}
\end{eqnarray}
where $\gamma_E$ is the usual Euler-Mascheroni constant. Similarly, the class of integrals for the non-planar families are:
\begin{eqnarray}
G^{N_{12}}_{a_1\cdots a_9}(x,s,\epsilon)&:=e^{2\gamma_E \epsilon} &\int \frac{d^dk_1}{i\pi^{d/2}}\frac{d^dk_2}{i\pi^{d/2}}
\frac{1}{k_1^{2a_1} (k_1 + x p_1)^{2a_2} (k_1 + x p_{12})^{2a_3} (k_1 + p_{123})^{2a_4}} \label{eq:N12x}\\
&\times& \frac{1}{k_2^{2a_5} (k_2 - x p_1)^{2a_6} (k_2 - p_{123})^{2a_7} 
(k_1 + k_2 + x p_2)^{2a_8} (k_1 + k_2)^{2a_9}}, \nonumber
\end{eqnarray}
\begin{eqnarray}
G^{N_{13}}_{a_1\cdots a_9}(x,s,\epsilon)&:=e^{2\gamma_E \epsilon} &\int \frac{d^dk_1}{i\pi^{d/2}}\frac{d^dk_2}{i\pi^{d/2}}
\frac{1}{k_1^{2a_1} (k_1 + x p_1)^{2a_2} (k_1 + x p_{12})^{2a_3} (k_1 + p_{123})^{2a_4}} \label{eq:N13x}\\
&\times& \frac{1}{k_2^{2a_5} (k_2 - x p_{12})^{2a_6} (k_2 - p_{123})^{2a_7} 
(k_1 + k_2 + x p_1)^{2a_8} (k_1 + k_2)^{2a_9}}, \nonumber
\end{eqnarray}
\begin{eqnarray}
G^{N_{34}}_{a_1\cdots a_9}(x,s,\epsilon)&:=e^{2\gamma_E \epsilon} &\hspace{-0.1cm}\int \frac{d^dk_1}{i\pi^{d/2}}\frac{d^dk_2}{i\pi^{d/2}}
\frac{1}{k_1^{2a_1} (k_1 + x p_1)^{2a_2} (k_1 + x p_{12})^{2a_3}
 (k_1 + p_{123})^{2a_4}} \label{eq:N34x}\\
&\times& \hspace{-0.3cm}\frac{1}{k_2^{2a_5} (k_2 - x p_1)^{2a_6} (k_2 - p_{123})^{2a_7} 
(k_1 + k_2 + x p_{12} - p_{123})^{2a_8} (k_1 + k_2)^{2a_9}}. \nonumber
\end{eqnarray}

Using the notation given in Eq. (\ref{eq:P12x}), (\ref{eq:P13x}) and (\ref{eq:P23x}) for the class of planar loop integrals, the indices $a_1\cdots a_9$ for the list of MI in the planar families are as follows\footnote{The letter m is used here as well as in the ancillary files to indicate the index -1.}:
{\footnotesize
\begin{eqnarray}
P_{12}: \hspace{-0.1cm} &\{010000011,001010001,001000011,100000011,101010010,101010100,101000110,010010101, \nonumber\\
&101000011,101000012,100000111,100000112,001010011,001010012,010000111,010010011, \nonumber\\
&101010110,111000011,101000111,101010011,011010011,011010012,110000111,110000112, \nonumber\\
&010010111,010010112,111010011,111000111,111010111,111\text{m}10111,11101\text{m}111\}, \label{eq:MIP12}
\end{eqnarray}
\begin{eqnarray}
P_{13}: \hspace{-0.1cm} &\{000110001,001000011,001010001,001101010,001110010,010000011,010101010,010110010, \nonumber\\
&001001011,001010011,001010012,001011011,001101001,001101011,001110001,001110002, \nonumber\\
&001110011,001111001,001111011,001211001,010010011,010110001,010110011,011010011, \nonumber\\
&011010021,011110001,011110011,011111011,\text{m}11111011\}, \label{eq:MIP13}
\end{eqnarray}
\begin{eqnarray}
P_{23}: \hspace{-0.1cm} &\{001010001,001010011,010000011,010000101,010010011,010010101,010010111,011000011, \nonumber\\
&011010001,011010010,011010011,011010012,011010100,011010101,011010111,011020011, \nonumber\\
&012010011,021010011,100000011,101000011,101010010,101010011,101010100,110000111, \nonumber\\
&111000011,111010011,111010111,111\text{m}10111\} \label{eq:MIP23}.
\end{eqnarray}
}
Most of the above planar MI also appear in the non-planar families, whose list of MI have the following indices $a_1\cdots a_9$ in terms of the notation of Eq. (\ref{eq:N12x}), (\ref{eq:N13x}) and (\ref{eq:N34x}):
{\footnotesize
\begin{eqnarray}
N_{12}: \hspace{-0.1cm} &\{100001010,000110010,000110001,000101010,000101001,101010010,100110010,100101020, \nonumber\\
&100101010,100101001,001110010,001110002,001110001,001101001,101110020,101110010, \nonumber\\
&101101002,101101001,100111020,100111010,100102011,100101011,001120011,001111002, \nonumber\\
&001111001,001110011,000111011,101011011,100111011,1\text{m}0111011,001111011,0\text{m}1111011, \nonumber\\
&101111011,1\text{m}1111011,1\text{m}1111\text{m}11\}, \label{eq:MIN12}
\end{eqnarray}
\begin{eqnarray}
N_{13}: \hspace{-0.1cm} &\{010000110,000110010,001000101,001000110,001010001,010110100,001110100,001010102, \nonumber\\
&001110002,000110110,001010101,001010110,001100110,001110001,001110010,010100110, \nonumber\\
&010110101,002010111,001120011,001210110,011010102,001110120,001010111,001110210, \nonumber\\
&001110011,001110101,001110110,002110110,011000111,011010101,011100110,011110001, \nonumber\\
&011110110,\text{m}11010111,010110111,\text{m}01110111,0\text{m}1110111,00111\text{m}111,001110111,011010111, \nonumber\\
&011110101,011110111,\text{m}11110111\}, \label{eq:MIN13}
\end{eqnarray}
\begin{eqnarray}
N_{34}: \hspace{-0.1cm} &\{001001010,001010010,010010010,100000110,100010010,000010111,010010110,001010102, \nonumber\\
&001010101,010010101,001020011,010000111,001010011,010010011,101010020,101010010, \nonumber\\
&101010100,101000011,110010120,110010110,010010112,010010121,010010111,010020111, \nonumber\\
&020010111,011010102,001010111,011010101,110000211,011020011,110000111,011010011, \nonumber\\
&111000101,111010010,101010101,101010011,111010110,111010101,101010111,11\text{m}010111, \nonumber\\
&110\text{m}10111,11001\text{m}111,110010111,\text{m}11010111,011\text{m}10111,01101\text{m}111,011010111,111000111, \nonumber\\
&111010011,111010111,111\text{m}10111\} \label{eq:MIN34}.
\end{eqnarray}
}

Our integrals (\ref{eq:P12x}) $-$ (\ref{eq:N34x}) are complex functions of $x,s_{12},s_{23}$ and $q$.
%that are defined on an open set of the complex plane. 
However, for phenomenological calculations of the process (\ref{eq:ppVV}) only the {\it physical region} in phase space is relevant, which is expressed in terms of the Mandelstam and mass variables as:
\begin{gather}
S>\left(\sqrt{M_3^2}+\sqrt{M_4^2}\right)^2, \ \ \ T<0, \ \ \ U<0, \nonumber\\
M_3^2>0, \ \ \ M_4^2>0, \ \ \ q_{\perp}^2=\frac{TU-M_3^2M_4^2}{S}>0, \label{eq:physreg}
\end{gather}
where $q_{\perp}^2$ is the transverse momentum squared of each massive particle in the centre of mass frame. In terms of our variables the physical region is therefore:
\begin{equation}
x>1, \ \ \ \frac{q-s_{12}}{s_{23}} > 1, \ \ \ xs_{12} > q, \ \ \ q>0. \label{eq:physregx}
\end{equation}
The second inequality in (\ref{eq:physregx}) causes a branching of the $s_{23}$ variable in the physical region:
\begin{equation}
  x > 1, \ \ \ \begin{cases}
    s_{23} < 0, \ \ \ s_{12} + s_{23} > q, \ \ \ q > 0 & \\
    s_{23} > 0, \ \ \ s_{12} + s_{23} < q, \ \ \ s_{12} > q/x.  & 
  \end{cases}
\end{equation}
%We note that the boundary terms of our MI are evaluated at $x=0$ as explained in the previous Section. 
In the next Section we will discuss our results for the MI and describe how they can be analytically continued to the physical region  (\ref{eq:physregx}).

\section{Results for Master Integrals}

With the chosen $x$-parametrization, the solutions of the DE are all expressed in terms of GPs (\ref{eq:GPs}). The set of all indices $I(P_i)$ of the GPs appearing in the MI of each planar family is:
\begin{gather}
I(P_{12})=\left\{0,1,\frac{q}{s_{12}},\frac{s_{12}}{q}, \frac{q}{q - s_{23}}, 1-\frac{s_{23}}{q}, 1+\frac{s_{23}}{s_{12}} ,\frac{s_{12}}{s_{12} + s_{23}}\right\}, \nonumber\\
I(P_{13})=\left\{0, 1, \frac{q}{s_{12}}, \frac{s_{12} + s_{23}}{s_{12}}, \frac{q}{q - s_{23}}, \xi_-, \xi_+, \frac{q (q - s_{23})}{q^2 - (q + s_{12}) s_{23}}\right\}, 
\label{eq:indicesP}\\
I(P_{23})=\left\{0, 1, \frac{q}{s_{12}}, 1+\frac{ s_{23}}{s_{12}}, \frac{q}{q - s_{23}}, \frac{q}{s_{12}+ s_{23}}, \frac{q - s_{23}}{s_{12}}\right\},\nonumber\\
%I(P_{23})=I(P_{12})\cup \left\{\frac{q-s_{23}}{s_{12}},\frac{q}{s_{12} + s_{23}}\right\}, \label{eq:indicesP}\\
%I(P_{13})=I(P_{12})\cup \left\{\xi_+,\xi_-,
%\frac{q (q - s_{23})}{q^2-s_{23} \left(q+s_{12}\right)}\right\}, \nonumber\\
\xi_{\pm}=\frac{q s_{12} \pm \sqrt{q s_{12} s_{23} (-q + s_{12} + s_{23})}}{q s_{12} - s_{12} s_{23}}. \nonumber
\end{gather}
Many of the above indices also appear in the GPs of the non-planar families:
\begin{gather}
I(N_{12})=I(P_{23}), \nonumber\\
I(N_{34})=I(P_{12})\cup I(P_{23}) \cup 
\left\{  \frac{s_{12}}{q - s_{23}}, \frac{s_{12} + s_{23}}{q}, \frac{q^2 - q s_{23} - s_{12} s_{23}}{s_{12} (q - s_{23})}, \frac{s_{12}^2 + q s_{23} + s_{12} s_{23}}{s_{12} (s_{12}+ s_{23})} \right\},  \\
I(N_{13})=I(P_{23})\cup \left\{ \xi_-, \xi_+, 1 +  \frac{q}{s_{12}} + \frac{q}{-q + s_{23}}\right\}. \nonumber
\end{gather}

Note that the parameter $x$ is not part of the set of indices but appears in the argument of the GPs as we integrate the DE. This can be for example explicitly seen in the solution of the scalar double box of the family $P_{13}$:
\begin{gather*}
G^{P_{13}}_{011111011}(x,s,\epsilon)=\frac{A_3(\epsilon)}{x^2 s_{12} (-q+x(q-s_{23}))^2}\left\{\frac{-1}{2\epsilon^4}
+\frac{1}{\epsilon^3}\!\left(\!-GP\left(\frac{q}{s_{12}};x\right)\!+2 \ \! GP\!\left(\!\frac{q}{q-s_{23}};x\!\right) \right.\right. \nonumber\\
\left.+2 \ GP(0;x)-GP(1;x)+\log \left(-s_{12}\right)+\frac{9}{4}\right)+\frac{1}{4\epsilon^2}\left(18 \ GP\left(\frac{q}{s_{12}};x\right)
-36 \ GP\left(\frac{q}{q-s_{23}};x\right) \right. \nonumber\\
\left.-8 \ GP\left(0,\frac{q}{s_{12}};x\right)+16 \ GP\left(0,\frac{q}{q-s_{23}};x\right)+8 \ GP\left(\frac{s_{23}}{s_{12}}
+1,\frac{q}{q-s_{23}};x\right)+\cdots\right) \nonumber\\
+\frac{1}{\epsilon}\left(9\left(GP\left(0,\frac{q}{s_{12}};x\right)+GP(0,1;x)\right)-4\left(GP\left(0,0,\frac{q}{s_{12}};x\right)+GP(0,0,1;x)\right)+\cdots\right) \nonumber\\
\left. +6 \left(GP\left(0,0,1,\xi_-;x\right)+GP\left(0,0,1,\xi_+;x\right)\right)
\!-\!2 \ \! GP \! \left(0,0,\frac{q}{q-s_{23}},\frac{q \left(q-s_{23}\right)}{q^2-s_{23} \left(q+s_{12}\right)};x\right)
\! +\cdots\right\}.
\end{gather*}
The above MI is expressed in terms of a common factor $A_3(\epsilon)$
\begin{gather}
A_3(\epsilon)=-e^{2\gamma_E \epsilon}\frac{\Gamma(1-\epsilon)^3\Gamma(1+2\epsilon)}{\Gamma(3-3\epsilon)}. \label{eq:Pinv2}
\end{gather}
As one notices from the example $G^{P_{13}}_{011111011}$ above, the solutions are in general not uniform in the weight of the GPs~\cite{Henn:2013pwa}. The Goncharov polylogarithms can be numerically evaluated with the Ginac library~\cite{Vollinga:2004sn}. We have not tried to simplify our analytical expressions by the use of 
symbol and co-product techniques~\cite{Duhr:2011zq,Duhr:2012fh}, as we were mostly interested in showing the applicability of the SDE method, but 
this is expected to be straightforward. The full expressions for all MI are available in the attached ancillary files.

In order to compute the MI in arbitrary kinematics, especially in the physical region, the GPs have to be properly analytically continued. In general all variables, including the
momenta invariants $s_{ij}$ ($s_{12}$, $s_{23}$ and $q$ in the present study) and the parameter $x$, would acquire an infinitesimal imaginary part, $s_{ij}\to s_{ij}+i \delta_{s_{ij}} \eta $, $x\to x+i \delta_{x} \eta $, with
$\eta \to 0$. The parameters $\delta_{s_{ij}}$ and $\delta_x$ are determined as follows: the first class of constraints on the above-mentioned parameters originates form 
%in the first place 
the input data to the DE, namely the one-scale master integrals
that need to be properly defined in each kinematical region. For instance writing the two-point function $G^{P{12}}_{001000011}$ as
\[
G^{P{12}}_{001000011}\sim \left(-(-1 + x) (-q + s_{12} x\right))^{1-2\epsilon}\sim (1-x)^{1-2\epsilon}\left(1-x s_{12}/q\right)^{1-2\epsilon}(-q)^{1-2\epsilon}
\]
implies constraints on the imaginary parts of the quantities involved. 
The second class of constraints results from the second graph polynomial~\cite{Bogner:2010kv}, ${\cal F}$, which after expressed in terms
of $s_{ij}$ and $x$, should acquire a definite-negative imaginary 
part in the limit $\eta\to 0$. Combining these two classes of constraints on the  parameters $\delta_{s_{ij}}$ and $\delta_x$, the imaginary part of all the GPs involved is fixed and we have checked that the result for the MI is identical in the limit $\eta\to 0$ and moreover agrees with the one obtained by other calculations.

We have performed several numerical checks of all our calculations. 
The numerical results have been compared with those provided by the numerical code {\bf SecDec} \cite{Binoth:2000ps,Heinrich:2008si,Borowka:2012yc,Borowka:2013cma} in the Euclidean region and with analytic results 
presented in~\cite{Henn:2014lfa,Caola:2014lpa} 
for the physical region. In all cases we find perfect agreement and reference numerical results can be found in the ancillary files.

\section{Conclusions}

In this paper we calculated the complete set of two-loop MI with massless propagators contributing to diboson production at the LHC and performed an independent calculation of the MI presented in~\cite{Henn:2014lfa}. The MI were calculated with the SDE method proposed in~\cite{Papadopoulos:2014lla}. After choosing the $x$-parameterization as given in this paper, all MI are expressible in terms of GPs, {\it independently on the number of kinematical scales involved}, and moreover the boundary terms are directly captured by integrating the singularities at the boundary point $x=0$. This seems to be the case for all MI calculated with the SDE method, which makes it an efficient and practical tool for the calculation of MI since it does not require the independent derivation of limits of those MI. We have shown that one $x$-parameterization of the external momenta is enough to parametrize both the planar as well as non-planar integrals and express them in terms of GPs with the argument $x$. The indices of the GPs are functions of the other three parameters that are linearly related to the scalar products of the external momenta when $x\rightarrow 1$.

%The phase space point at the boundary $x=0$ lies in the Euclidean region and 
We have also discussed how to analytically continue the solutions for the MI into the physical region. All our results have been numerically compared with {\bf Secdec} in the Euclidean
region and with the analytical results of~\cite{Henn:2014lfa,Caola:2014lpa} in the physical region for various phase space points and perfect agreement has been found. All MI are attached as ancillary files to this paper together with reference  
numerical results for the MI with 7 denominators. Our results may also be used to calculate virtual two-loop QCD corrections contributing to processes at the LHC such as $WZ$, $V\gamma^*$, $ZH$, or to calculate sub-diagrams of three-loop Feynman diagrams.

There are still several open issues which need further study. In particular it will be useful to find all $x$-parameterizations 
%will lead to DE that when integrated results in GPs. One may ask if it is possible to find a correct $x$-parameterization 
of the external momenta that {\it a priori} lead to DE directly expressible in terms of Goncharov polylogarithms.
%, in other words before having to calculate the integrating factors. 
Moreover, as the DE are directly related to the IBP identities, it is a very intriguing question if such a parameterization can be determined by studying the IBP identities themselves. 
Another subject of future research is the extension of the SDE method to integrals with massive internal lines and more external legs.

\subsection*{Acknowledgements}

This research was supported by the Research Funding Program ARISTEIA, HOCTools (co-financed by the European Union (European Social Fund ESF) and Greek national funds through the Operational Program "Education and Lifelong Learning" of the National Strategic Reference Framework (NSRF)). We gratefully acknowledge fruitful discussions with S. Borowka, T. Hann, G. Heinrich, V. Yundin, J. Henn, A. V. Smirnov M. Czakon and S. Weinzierl during various stages of this project.

\appendix
\section{Expansion by regions}
\label{expbyreg}

In this appendix we present a few examples to demonstrate that the $x\to 0$ limit of the MI is captured by the one-scale integrals involved in the DE.
We consider all scales made by scalar products of the external momenta as the hard scale, i.e. $p_{123}^2\sim p_{12}^2\sim p_{123}.p_{12}$.
\\[12pt]
\noindent $\bullet$ Planar double box: $I:=G^{P13}_{011111011}$\\
The integral is given by:
\begin{equation}
I=\int \frac{d^dk_1}{i\pi^{d/2}}\frac{d^dk_2}{i\pi^{d/2}}
\frac{1}{(k_1 + x p_1)^2 (k_1 + x p_{12})^2 (k_1 + p_{123})^2 k_2^2 (k_2 - x p_1)^2 (k_2 - p_{123})^2 (k_1 + k_2)^2}.
\end{equation}

The integrating factor of the DE of $I$ behaves as $M(x)\sim x^2$ as $x\rightarrow 0$ and we are therefore interested in the behaviour
%\footnote{Note that any possible overlapping regions are further suppressed and may therefore be neglected.}
$\lim_{x\rightarrow 0}MI(x)=\lim_{x\rightarrow 0} x^2I=\lim_{x\rightarrow 0}x^2\sum_{R\in\mathcal{R}}I^{(R)}$, where ${\cal R}$ denotes the set of all regions contributing. 
Performing the expansions for all the regions given above, it can be shown by power counting that we have a vanishing limit $\lim_{x\rightarrow 0}(x^2I^{(R)})=0$ for all regions except the purely soft region $R=(\mathcal{S},\mathcal{S})$,
which is defined by $\mathcal{S}\:=k\sim x p_{123}$ ($k$ is either $k_1$ or $k_2$). Therefore, the behaviour of $MI(x)$ as $x\rightarrow 0$ is fully determined by the soft region:
\begin{gather}
\lim_{x\rightarrow 0}(x^2I^{(\mathcal{S},\mathcal{S})})=\lim_{x\rightarrow 0}x^2 \int \frac{d^dk_1}{i\pi^{d/2}}\frac{d^dk_2}{i\pi^{d/2}}
\frac{1}{(k_1 + x p_1)^2 (k_1 + x p_{12})^2 k_2^2 (k_2 - x p_1)^2 (k_1 + k_2)^2(p_{123}^2)^2} \nonumber\\
\stackrel{IBP}{=}\frac{(-1+2\epsilon)(-2+3\epsilon)(-1+3\epsilon)}{2\epsilon^3s_{12}^2q^2}\lim_{x\rightarrow 0}x^{-2}G^{P13}_{001010001}=\frac{(2+9(1-\epsilon)\epsilon)(-s_{12})^{-2\epsilon}A_3(\epsilon)}{4\epsilon^4q^2s_{12}x^{4\epsilon}}. \label{eq:P13}
\end{gather}
On the other hand, the singular part of the DE of $MI(x)$ is exactly given by the above single scale 2-point MI, namely  $G^{P13}_{001010001}$. Integrating the singular term of the DE we obtain the result shown in Eq. (\ref{eq:P13}), as can be seen from the ancillary files. 
We conclude that the DE correctly captures the single-scale behaviour $\sim G^{P13}_{001010001}$ as $x\rightarrow 0$ via the re-summed term.
\\[12pt]
\noindent $\bullet$ Non-planar double box: $G^{N12}_{101111011}$\\
The integral equals:
\begin{equation}
I=\int \frac{d^dk_1}{i\pi^{d/2}}\frac{d^dk_2}{i\pi^{d/2}}
\frac{1}{k_1^2 (k_1 + x p_{12})^2 (k_1 + p_{123})^2 k_2^2 (k_2 - x p_1)^2 (k_1 + k_2 + x p_2)^2 (k_1 + k_2)^2}.
\end{equation}

The integrating factor of the DE of $I$ behaves as $M(x)\sim x^{4+2\epsilon}$ as $x\rightarrow 0$ and we are therefore interested in the behaviour $\lim_{x\rightarrow 0}MI(x)=\lim_{x\rightarrow 0}x^{4+2\epsilon}\sum_{R\in\mathcal{R}}I^{(R)}$. 
Performing the expansions for all the regions given above, it can be shown by power counting that $\lim_{x\rightarrow 0}(x^{4+2\epsilon}I^{(R)})=0$ for all regions except the purely soft region $R=(\mathcal{S},\mathcal{S})$. 
Therefore, the behaviour of $MI(x)$ as $x\rightarrow 0$ is fully determined by the soft region:
\begin{gather}
\lim_{x\rightarrow 0}(x^{4+2\epsilon}I^{(\mathcal{S},\mathcal{S})})=\lim_{x\rightarrow 0}\frac{x^{4+2\epsilon}}{q}G^{N12}_{101011011}=A_3(\epsilon) F(\epsilon)\frac{(-s_{12})^{-2\epsilon}x^{-2\epsilon}}{q s_{12}^2}, \label{eq:N12}
\end{gather}
where $F$ is an $\epsilon$-dependent function given by 
\[ 
F(\epsilon)= 
\frac{
36 - 234\epsilon - 6\epsilon^2 (-81 + 5\pi^2) + 3\epsilon^3(-108 + 65\pi^2 - 204 Z_3) - \epsilon^4(405 \pi^2 + 17\pi^4 - 3978 Z_3) 
}{18 \epsilon^4 (-1 + 2 \epsilon)}
\]
The singular part of the DE of $MI(x)$ is again exactly given by the above single scale 3-point MI $G^{N12}_{101011011}$. 
Integrating the singular term of the DE results in Eq. (\ref{eq:N12}) as can be seen from the ancillary files. 
We again conclude that the DE correctly captures the single-scale behaviour $\sim G^{N12}_{101011011}$ as $x\rightarrow 0$ via the re-summed term.
\\[12pt]
\noindent $\bullet$  2-loop triangle: $G^{N34}_{101010010}$\\
The integral equals:
\begin{equation}
I(x)=\int \frac{d^dk_1}{i\pi^{d/2}}\frac{d^dk_2}{i\pi^{d/2}}
\frac{1}{k_1^2 (k_1 + x p_{12})^2 k_2^2 (k_1 + k_2 - p_{123} + x p_{12})^2} \label{eq:N34101010010}.
\end{equation}

The integrating factor of the DE of $I$ is $M(x)= x^{3\epsilon}$  and we are  interested in the behaviour $\lim_{x\rightarrow 0}M(x) I(x)=\lim_{x\rightarrow 0}x^{3\epsilon}\sum_{R\in\mathcal{R}}I^{(R)}$. In this case, the regions contribute to the behaviour of $MI(x)$ as $x\rightarrow 0$:
\begin{gather}
\lim_{x\rightarrow 0}(x^{3\epsilon}I^{(\mathcal{H},\mathcal{H})})=x^{3\epsilon}G^{N34}_{002010010}\stackrel{IBP}{=}\frac{x^{3\epsilon}(1-2\epsilon)(2-3\epsilon)}{\epsilon q}G^{N34}_{001010010} \nonumber\\
=\frac{(2-3\epsilon)A_3(\epsilon)(-q)^{-2\epsilon}}{2\epsilon^2}x^{3\epsilon}, \nonumber\\
\lim_{x\rightarrow 0}(x^{3\epsilon}I^{(\mathcal{S},\mathcal{H})})=x^{3\epsilon}G^{N34}_{101010100}=\frac{(-q)^{-\epsilon}(-s_{12})^{-\epsilon}c_{\Gamma}(\epsilon)^2}{(1-2\epsilon)^2\epsilon^2}x^{\epsilon}. \label{eq:N34}
\end{gather}
where $\mathcal{H}:=k\sim p_{123}$. The DE of $M(x)I(x)$ and its coupled MI $I_2(x):=G^{N34}_{101010020}$ equals:
\begin{gather}
\partial_x (M(x)I(x))=\frac{(2-3\epsilon)G^{N34}_{100010010}}{q(1-x)x^{1-3\epsilon}(1-xs_{12}/q)}-\frac{q\epsilon \left(M_2(x) I_2(x)\right)}{(1-2\epsilon)(1-x)^{2\epsilon}
x^{2-4\epsilon}(1-xs_{12}/q)^{2\epsilon}},\\
\partial_x (M_2(x)I_2)=\frac{(1-2\epsilon)(2-3\epsilon)(1-3\epsilon)G^{N34}_{001010010}}{\epsilon q^2(1-x)^{1-2\epsilon}x^{\epsilon}(1-x s_{12}/q)^{1-2\epsilon}}
-\frac{(1-2\epsilon)(1-3\epsilon)\left(M(x) I(x)\right)}{q (1-x)^{1-2\epsilon}x^{4\epsilon}(1-x s_{12}/q)^{1-2\epsilon}}, \nonumber
\end{gather}
where $M_2(x)= (1-x)^{2\epsilon}x^{1-\epsilon}(1-xs_{12}/q)^{2\epsilon}$ is the integrating factor of $I_2(x)$. 
Moreover

\[ G^{N34}_{100010010} =-\frac{ A_3(\epsilon) }{ 2 \epsilon (1-2\epsilon) }(-q)^{1-2\epsilon}(1-x)^{1-2\epsilon}(1-x s_{12}/q)^{1-2\epsilon} 
\]
and 
\[ G^{N34}_{001010010} =-\frac{ A_3(\epsilon)  }{ 2 \epsilon (1-2\epsilon) }(-q)^{1-2\epsilon} .
\]

We observe that the singularity in the inhomogeneous term of $\partial_x (MI)$ does not equal the differential of the regions $x^{3\epsilon}I^{(\mathcal{H},\mathcal{H})}$ and $x^{3\epsilon}I^{(\mathcal{S},\mathcal{H})}$ as given in (\ref{eq:N34}). In fact, the remaining singularity in the DE of $M(x)I(x)$ lies in the homogeneous term $I_2(x)$. This can be seen by considering the behaviour of $I_2$ as $x\rightarrow 0$, which gets contributions from the same regions as in Eq. (\ref{eq:N34}):
\begin{gather}
\lim_{x\rightarrow 0}I_2^{(\mathcal{H},\mathcal{H})}=G^{N34}_{002010020}\stackrel{IBP}{=}\frac{2(1-2\epsilon)(2-3\epsilon)(-1+3\epsilon)}{\epsilon q^2}G^{N34}_{001010010} \nonumber\\
=\frac{(2-3\epsilon)(1-3\epsilon)A_3(\epsilon)(-q)^{-1-2\epsilon}}{\epsilon^2}, \nonumber\\
\lim_{x\rightarrow 0}I_2^{(\mathcal{S},\mathcal{H})}=G^{N34}_{101020100}\stackrel{IBP}{=}\frac{(1-2\epsilon)}{-q}G^{N34}_{101010100}=\frac{(-q)^{-1-\epsilon}(-s_{12})^{-\epsilon}c_{\Gamma}(\epsilon)^2}{(1-2\epsilon)\epsilon^2}x^{-2\epsilon}. \label{eq:N34dot}
\end{gather}
Furthermore, the behaviour of the region $(\mathcal{S},\mathcal{H})$ of $I$, which scales as $G^{N34}_{101010100}$, does not appear at all in the inhomogeneous part of $\partial_x (MI)$ and lies instead fully in the homogeneous term $I_2$. The reason for this is that the MI $G^{N34}_{101010100}$ cannot be reproduced from the graph $G^{N34}_{101010010}$ by pinching internal lines. 
%As the DE of $M_2(x)I_2$ does not contain any singularities, we cannot capture the behaviour of either regions by simply integrating the singular parts of the inhomogeneous terms.

Let us consider now instead the integral $I(1/x)$, found from Eq. (\ref{eq:N34101010010}) by $x\rightarrow 1/x$. We perform some shifts and rescalings to get the integral into the form:
\begin{equation*}
I(1/x)=x^{4\epsilon}\int \frac{d^dk_1}{i\pi^{d/2}}\frac{d^dk_2}{i\pi^{d/2}}
\frac{1}{k_1^2 (k_1 - p_{12})^2 (k_2 + x p_{123})^2 (k_1 + k_2)^2}.
\end{equation*}
The regions in terms of the above loop momenta $k_1,k_2$ are given by:
\begin{equation}
\mathcal{R}=\{(\mathcal{H},\mathcal{H}),(\mathcal{S},\mathcal{S})\}.
\end{equation}
The integrating factor of the DE of $I(1/x)$ behaves as $M(1/x)\sim x^{-3\epsilon}$ as $x\rightarrow 0$ and we are therefore interested in the behaviour $\lim_{x\rightarrow 0}MI(1/x)=\lim_{x\rightarrow 0}x^{-3\epsilon}\sum_{R\in\mathcal{R}}I^{(R)}$. Performing the expansion, it can be shown by power counting that $\lim_{x\rightarrow 0}(x^{-3\epsilon}I^{(\mathcal{S},\mathcal{S})})=0$. Therefore, the behaviour of $MI(1/x)$ as $x\rightarrow 0$ is fully determined by the hard region:
\begin{gather}
\lim_{x\rightarrow 0}(x^{-3\epsilon}I^{(\mathcal{H},\mathcal{H})})=x^{-3\epsilon}\int \frac{d^dk_1}{i\pi^{d/2}}\frac{d^dk_2}{i\pi^{d/2}}
\frac{1}{k_1^2 (k_1 - p_{12})^2 k_2^2 (k_1 + k_2)^2} \nonumber\\
\stackrel{IBP}{=}\frac{x^{2-3\epsilon}(-2+3\epsilon)}{s_{12}\epsilon}G^{N34}_{100000110}(1/x)=\lim_{x\rightarrow 0}\frac{x^{2-3\epsilon}(-2+3\epsilon)}{s_{12}\epsilon}G^{N34}_{100010010}(1/x) \nonumber\\
=-\frac{(2-3\epsilon)A_3(\epsilon)(-s_{12})^{-2\epsilon}x^{\epsilon}}{2\epsilon^2(1-2\epsilon)}. \label{eq:N34dx}
\end{gather}
with
\[ G^{N34}_{100010010}(1/x) =-\frac{ A_3(\epsilon) }{ 2 \epsilon (1-2\epsilon) }(-s_{12})^{1-2\epsilon}(1-x)^{1-2\epsilon}(1-x q/s_{12})^{1-2\epsilon} x^{-2+4\epsilon}.
\]
Contrary to the case of $I(x)$, in this case the resulting integrals $G^{N34}_{100000110}(1/x)$ and $G^{N34}_{100010010}(1/x)$ can be found from $I(1/x)$ by pinching internal lines. In fact, the singular part in the inhomogeneous term\footnote{It can easily be checked that in this case the homogeneous term $I_2(1/x)$ has no singularities since $I_2(1/x)\sim x^2$.} of the DE of $MI(1/x)$ is exactly given by the above single scale 2-point MI $G^{N34}_{100010010}$. Integrating this singular inhomogeneous term of the DE of $I(1/x)$ results in Eq. (\ref{eq:N34dx}), after transforming back $x\rightarrow 1/x$. We conclude that the inhomogeneous term of the DE of $I(1/x)$ correctly captures the single-scale behaviour $\sim G^{N34}_{100010010}(1/x)$ as $x\rightarrow 0$. The integral $I(1/x)$ can thus be found by solving as usual the DE of $I(1/x)$, without any other single scale integral input, and then $I(x)$ is found by inverting back $x\rightarrow 1/x$.

\bibliographystyle{JHEP}
\bibliography{doublebox_paper-2}

\end{document}